\documentclass{article}

\begin{document}

%
%
\def\REFERENCES
     {\countdef\refno=14
     \refno = 0\vskip 24pt
     {\centerline{\bf REFERENCES}\par\nobreak}
      \parindent=0pt
      \vskip 12pt\frenchspacing}
\def\ref#1#2#3
   {\advance\refno by 1
    {\rm[\number\refno] }\rm #1\ {\sl #2}
    \rm #3\par}

\topskip=0.40truecm

\baselineskip=18pt
\hsize=16truecm
\vsize=22truecm

\hoffset -2cm

\centerline{\bf ALGEBRAICALLY LINEARIZABLE DYNAMICAL SYSTEMS}

\bigskip\bigskip\bigskip

\centerline{\it R.Caseiro$^*$ and J.P. Fran\c{c}oise$^{**}$}

\bigskip

\centerline{$^*$Universidade de Coimbra}

\centerline{Departamento de Matem\'atica}

\centerline{3000 Coimbra, Portugal.}

\centerline{and}

\centerline{$^{**}$Universit\'e de Paris 6, UFR 920, tour 45-46, 4 place
Jussieu, B.P. 172,}

\centerline{Equipe "G\'eom\'etrie Diff\'erentielle, Syst\`emes Dynamiques
et Applications"}

\centerline{75252 Paris, France.}

\bigskip
\bigskip \bigskip\bigskip
\bigskip\bigskip


\noindent
{\it Summary}

The  main result of this paper is the evidence of an explicit
linearization of dynamical systems of Ruijsenaars-Schneider
(RS) type and of the perturbations introduced by F. Calogero of
these systems with all orbits periodic of same period. Several
other systems share the existence of this explicit linearization,
among them, the Calogero-Moser system (with and without external
potential) and the Calogero-Sutherland system. This
explicit linearization is compared with the notion of maximal
superintegrability which has been discussed in several
articles (to quote few of them, Hietarinta [12], Henon
[11], Harnad-Winternitz [10], 
S. Wojchiechowsky [15]).
\vskip 24pt

\noindent
{\it Short title}: Superintegrability


\vfill\eject
\noindent
{\bf  Introduction}

 Let $H: V^{2m}\rightarrow R$ be a Hamiltonian system defined
on a symplectic manifold $V^{2m}$, of dimension $2m$,
equipped with a symplectic form $\omega$
of dimension $2m$. Recall that $H$ is said to be integrable in
Arnol'd-Liouville sense if $H$ displays $m$ generically
independent first integrals (one of these maybe the Hamiltonian
itself) which are in involution for the Poisson bracket associated
with the symplectic form $\omega$. A vector field $X$ on a
manifold $V$ of dimension $n$ defines a flow and a dynamical
system. 
The vector field (not necessarily Hamiltonian) is classically said to
be maximally superintegrable if it has $n-1$ generically
independent globally defined first integrals $f_{1},...,f_{n-1}$. 
The orbits
of $X$ are then contained in the connected components of the
common level sets of the functions $f_{i}, i=1,...,n-1$. Some
Hamiltonian systems are known to be maximally superintegrable and
so they display $2m-1$ first integrals. This is so for instance of
the rational Calogero-Moser system, the Kepler problem, the
isotropic oscillator (cf. [1], [10], [11], [12], [15]).
Recently,
this specific class of Hamiltonian systems has deserved interest
in several articles (cf. for instance, [10]). In this
article the definition of algebraic linearization is proposed
in a slightly broader sense (also more precise sense).


\vskip 8pt
{\bf Definition}

A differential system is algebraically (resp. analytically) linearizable
if there are $n$ globally defined functions (rational, resp.
meromorphic) which are generically independent so that the time
evolution of the flow expressed in these functions is linear (in time)
and algebraic in the initial coordinates.

\

First purpose of this article is to prove that the
Hyperbolic and the Rational Ruijsenaars-Schneider systems are
algebraically linearizable. The perturbations recently considered
by Calogero [3], [4] of the Hyperbolic and Rational
Ruijsenaars-Schneider systems which display only periodic orbits
of the same period are algebraically linearizable as well.
This is proved rather easily using the Lax matrix first
introduced by Bruschi-Calogero [2] and the extra-equation which
implements the integrability of these systems first introduced in
the article [8].


\vskip 12pt
\noindent
{\bf I. Algebraic linearization of the rational
Calogero-Moser system and of the rational Calogero-Moser system with an
external
quadratic potential}

\


This first paragraph is devoted to the proof that the
rational Calogero-Moser system (with or without) external quadratic
potential is algebraically linearizable. The usefullness of
this (apparently) new notion is displayed on these classical
examples.
The rational Calogero-Moser system is represented by the Hamiltonian

$$H=(1/2)\sum_{i=1}^{m} y_{i}^{2} + g^{2}\sum_{i j}(x_{i}-x_{j})^{-2}
 \eqno(1.1)$$
where the constant $g$ is a parameter. J. Moser introduced the matrix
function:

$$L(x,y)/ L_{ij}= y_{i}\delta_{ij} + g{\rm i}(x_{i}-x_{j})^{-1}
(1-\delta_{ij}),\eqno(1.2)$$
and observed that the time evolution of this matrix function $L(x,y)$
along the flow is of Lax pair type:

$${\dot{L}}= [L,M]. \eqno(1.3)$$

        This Lax pair equation is supplemented with the equation:

$$\dot{X}=[X,M] + L, \eqno(1.4)$$
displayed by the diagonal matrix $X$:

$$X(x,y)/ X_{ij}=x_{i}\delta_{ij}. \eqno(1.5)$$

Following the classical approach, introduce the rational
functions:

$$F_{k}= tr(L^{k}). \eqno(1.6)$$

The Lax matrix equation yields:

$$\dot{F}_{k}=0. \eqno(1.7)$$

Introduce then the functions:

$$G_{k}= tr(XL^{k}), \eqno(1.8)$$
which undergo the time evolution:

$$\dot{G}_{k}= F_{k+1}. \eqno(1.9)$$

Clearly, the whole collection of the rational functions $F_{k}, G_{k}$
provide the algebraic linearization of the system.

\

Indeed, classical superintegrability can be recovered as follows:

$$ F_k \;(k=1,\ldots ,m) \;\;\; H_k=F_kG_k-F_{k+1}G_{k-1} \;
 (k=1, \ldots ,m-1)$$
provide $2m-1$ integrals of motion.


The next classical example to be considered is the
rational Calogero-Moser system with an external quadratic
potential. The system is described by the Hamiltonian:

$$H=(1/2)\sum_{i=1}^{m} y_{i}^{2} + g^{2}\sum_{i j}(x_{i}-x_{j})^{-2}
 +(\lambda^{2}/2)\sum_{i=1}^{m} y_{i}^{2}. \eqno(1.10)$$

The equations (1.3) and (1.4), with the same matrices $L$
 and $X$ get modified as follows:

$${\dot{L}}= [L,M]-{\lambda}^{2}X, \eqno(1.11a)$$

$$\dot{X}=[X,M] + L. \eqno(1.11b)$$

The classical approach consists in (cf [9]) introducing
matrices:

$$Z= L+{\rm i}\lambda X, \eqno(1.12a)$$

$$W= L-{\rm i}\lambda X. \eqno(1.12b)$$

These matrices undergo the time evolution:

$$\dot{Z}={\rm i}\lambda Z+[Z,M], \eqno(1.13a)$$

$$\dot{W}=-{\rm i}\lambda W+[W,M]. \eqno(1.13b)$$

It was then observed ([9]) that the matrix $P=ZW$ defines
Lax matrix for the system:

$$\dot{P}=[P,M]. \eqno(1.14)$$

        Here, we note that the functions:

$$F_{k}=tr(ZP^{k}), \eqno(1.15a)$$

$$G_{k}=tr(WP^{k}), \eqno(1.15b)$$
yield:

$$\dot{F}_{k}={\rm i}\lambda F_{k}, \eqno(1.16a)$$

$$\dot{G}_{k}={\rm i}\lambda G_{k}. \eqno(1.16b)$$

Thus these functions provide the algebraic
linearization of the system.


\


\

{\bf II. Algebraic linearization of the Calogero-Sutherland system}

\


 The Calogero-Sutherland system is defined by the Hamiltonian

$$H(x,y)=\frac{1}{2}\sum_{i=1}^{m}y_i^2 + \frac{g^2}{2}
\sum_{\stackrel{i,j=1}{i\neq j}}^{m}\sinh^{-2}(x_i-x_j) \eqno(2.1)$$
and has a Lax pair  
${\displaystyle \dot{L}=[L,M]_-}$ with Lax matrix

$$L_{ij}=y_i\delta_{ij}+\frac{\sqrt{-1}g}{sinh(x_i-x_j)}(1-\delta_{ij}). 
\eqno(2.2)$$

Defining the matrix $X$ by

$$X_{ij}=\exp (2x_i)\delta_{ij}, \eqno(2.3)$$
we get the dynamical equation

$$\dot{X}=[X,L]_++[X,M]_-. \eqno(2.4)$$
Above and throughout of course $[A,B]_{-}\equiv{AB - BA}$ and
$[A,B]_{+}\equiv AB + BA$.

Consider the functions

$$F_k=Tr(L^k), \;\;\; k=1,\ldots,m \eqno(2.5a)$$

$$G_k=Tr(XL^k), \;\;\; k=1,\ldots.m \eqno(2.5b)$$

The functions $F_{k}$ are first integrals of the dynamical system
defined by (2.1). Newton's formulae relate these constant of
motion with the coefficients $A_{0},..., A_{n-1}$ of the
characteristic polynomial of the matrix $L$. The theorem yields:

$$L^{n}= A_{n-1}L^{n-1}+A_{n-2}L^{n-2}+...+A_{0}I. \eqno(2.6)$$

\

{\bf Theorem II-1}

The functions $G_k$ undergo a linear evolution under the time 
evolution of the system defined by (2.1).

\

{\bf Proof:}

Once (L,M) is a Lax pair of the system and $X$ satisfyies (2.3),

$$\dot{G}_{k}= tr(XL^{k+1})=G_{k+1}. \eqno(2.7)$$

Thus, the vector $G= (G_{0},..., G_{n-1})$ displays the time
evolution:

$$\dot{G}= A G, \eqno(2.8)$$
where the matrix $A$ is with coefficients first integrals of the
differential system:

$$A_{ij}=\delta_{i+1,j}+A_{j-1}\delta_{in}. \eqno(2.9) $$

\

So, the Calogero-Sutherland system is algebraically linearizable.

\vskip 12pt
\noindent
{\bf III. Algebraic linearization of Hyperbolic and Rational
Ruijsenaars-Schneider systems}

        The dynamical systems of Ruijsenaars-Schneider (RS) type
 characterized by the equations of motion

$$\ddot{z}_j = \sum^n_{k=1 , k\not= j} \dot{z}_j \dot{z}_k f (z_j
- z_k), \,\,\, j=1,...,n,\eqno(3.1)$$
are ``integrable'' or ``solvable'' [4], if

$$f (z) = 2/z~~~~~~ {\it ``case (i)"},\eqno(3.2a)$$

$$f (z) = 2 / [z (1 + r^2 z^2)]~~~ {\it ``case (ii)"},\eqno(3.2b)$$

$$f(z) = 2 a {\rm cotgh} (az)~~~ {\it ``case (iii)"},\eqno(3.2c)$$

$$f(z) = 2a / {\rm sinh} (az)~~~{\it ``case (iv)"},\eqno(3.2d)$$

$$f (z) = 2a {\rm cotgh} (az) / [1 + r^2 {\rm sinh}^2 (az)]~~~ 
``{\it case (v)"},\eqno(3.2e)$$

$$f(z) = - a {\cal {P}}' (az)/ [{\cal {P}} (az) - {\cal {P}} (ab)]~~~
{\it ``case (vi)"}.\eqno(3.2f)$$

Of course the solutions $z_j (t)$ of (3.1) move in the complex plane; and
indeed all the
constants appearing in (3.2), namely $r$, $a$ and $b$, as well as the
constants $\omega$
and $\omega'$ implicit in the definition of the Weierstrass function
${\cal {P}} (z) \equiv
{\cal {P}} (z \vert \omega , \omega')$, might be complex.

Indeed, the main contribution of this paper is to solve explicitly
several systems following a scheme which may be of broader interest.

     The starting point of the analysis is the observation [2] that
(3.1) with (3.2e)
 is equivalent to the following ``Lax-type'' ({\it n}$\times${\it
n})-matrix equation:

$$\dot{L} = [L,M]_{-},\eqno(3.3)$$
with
$$L_{jk} = \delta_{jk} \dot{z}_j + (1 - \delta_{jk}) (\dot{z}_j
\dot{z}_k)^{1/2} \alpha
(z_j - z_k),\eqno(3.4)$$

$$M_{jk} = \delta_{jk} \sum^n_{m=1, m\not= j} \dot{z}_m \beta (z_j - z_m)
+ (1 -
\delta_{jk}) (\dot{z}_j \dot{z}_k)^{1/2} \gamma (z_j - z_k),\eqno(3.5)$$
and

$$\alpha (z) = \sinh (a \mu) / \sinh [a(z + \mu)],\eqno(3.6a)$$
$$\beta (z) = - a {\rm cotgh} (a\mu) / [1 + r^2 \sinh^2 (az)],\eqno(3.6b)$$
$$\gamma (z) = - a {\rm cotgh} (az) \alpha (z),\eqno(3.6c)$$
where
$$\sinh (a \mu) = {\rm i}/r.\eqno(3.7d)$$

It was furthermore recently noted [8] that the diagonal matrix
$$X(t) = {\rm diag} \{\exp [2a z_j (t)]\},\eqno(3.8)$$
undergoes the following time evolution:
$$\dot{X} = [X,M]_{-} + a [X, L]_{+}.\eqno(3.9)$$

 Let $F_{k}$ and $G_{k}$ be the functions defined as:

$$ F_{k}= tr(L^{k}), G_{k}= tr(XL^{k}). \eqno(3.10)$$

The functions $F_{k}$ are first integrals of the dynamical system
defined by (3.1). Newton's formulae relate these constant of
motion with the coefficients $A_{0},..., A_{n-1}$ of the
characteristic polynomial of the matrix $L$. The theorem yields:

$$L^{n}= A_{n-1}L^{n-1}+A_{n-2}L^{n-2}+...+A_{0}I, \eqno(3.11)$$

{\bf Theorem III-1}

        The functions $G_{k}$ undergo a linear evolution under the
time evolution of the system (3.1).

\

{\bf Proof:}

The equations (3.3) and (3.9) yield:

$$\dot{G}_{k}= 2atr(XL^{k+1})= 2aG_{k+1}.\eqno(3.12)$$

Thus,the vector $G= (G_{0},..., G_{n-1})$ displays the time
evolution:

$$\dot{G}= A G, \eqno(3.13)$$
where the matrix $A$ is with coefficients first integrals of the
differential system:

$$A_{ij}= 2a\delta_{i+1,j}+2a A_{j-1}\delta_{i,n}. \eqno(3.14) $$

F. Calogero introduced the following perturbation of the
trigonometric and rational Ruijsenaars-Schneider
systems characterized by the equations of motion:

$$\ddot{z}_j + {\rm i} \Omega \dot{z}_j
= \sum^n_{k=1 , k\not= j} \dot{z}_j \dot{z}_k f (z_j
- z_k), \,\,\, j=1,...,n.\eqno(3.15)$$
       
 F. Calogero made the remarkable conjecture [3], now proved in
the trigonometric and rational cases, that all the orbits of the
dynamical system defined by (3.15) are periodic of period
$\Omega$. The equations under consideration here are modified due
to the presence of the perturbation. The Lax equation (3.3) gets
modified into (cf. [8]):

$$\dot{L} = [L,M]_{-}+{\rm i}\Omega L,\eqno(3.16)$$
and the time evolution of the matrix $X$ is not modified.
        This yields new time evolution for the functions $F_{k}$ and
$G_{k}$:

$$\dot{F}_{k}= {\rm i}\Omega k F_{k} \eqno(3.17a).$$

$$\dot{G}_{k}= 2atr(XL^{k+1})+{\rm i}\Omega k tr(XL^{k})= 2aG_{k+1}
+{\rm i}\Omega k G_{k}\eqno(3.17b).$$


\vfill\eject
\REFERENCES

\ref{Barucchi, G., Regge, T.}{}{Conformal properties of a class of 
exactly solvable $N$-body problems in space dimension one. 
J. Math. Phys., {\bf 18}, n$^o$6, 1149-1153 (1977).}

\ref{Bruschi, M., Calogero, F.:}{}{ The Lax representation for an
integrable class of
relativistic dynamical systems. Commun. Math. Phys. {\bf 109}, 481-492
(1987).}

\ref{Calogero, F.:}{}{A class of integrable hamiltonian systems whose
solutions are (perhaps) all completely periodic. J.Math. Phys. {\bf 38},
5711-5719 (1997).}

\ref{Calogero, F.:}{}{ Tricks of the trade: relating and deriving 
solvable and integrable
dynamical systems. To appear in the Proceedings of the International
Workshop on
Calogero-Moser-Sutherland type models, held at the Centre de Recherches
Math\'ematiques de l'Universit\'e de Montr\'eal, Canada, in March 1997
(Springer, in press).}

\ref{Calogero, F.:}{}{ Motion of poles and zeros of nonlinear and linear
partial differential
equations and related many-body problems. Nuovo Cimento {\bf 43B}, 
177-241 (1978).}

\ref{Calogero, F.:}{}{ A solvable N-body problem in the plane. I. J. Math.
Phys. {\bf 37},
1735-1759 (1996).}

\ref{Calogero, F.:}{}{ Integrable and solvable many-body problems in the
plane via
complexification. J. Math. Phys., {\bf 39}, 5268-5291 (1998).}

\ref{Calogero, F., Fran\c{c}oise, J.-P.:}{}{Solution of certain integrable dynamical
systems of Ruijsenaars-Schneider type with completely periodic trajectories. To appear
}

\ref{Fran\c{c}oise, J.-P.:}{}{ Canonical partition functions of
Hamiltonian systems and the
stationary phase formula. Commun. Math. Phys. {\bf 117}, 37-47 (1988).}

\ref{Harnad, J., Winternitz, P.}{}{Harmonics on hyperspheres, 
separation of variables and the Bethe ansatz. Lett. Math. Phys. 
{\bf 33}, n$^o$ 1, 61-74 (1995).}

\ref{Henon, M.}{}{ Numerical exploration of Hamiltonian systems. 
Col: Chaotic behavior of deterministic systems. (Les Houches 1981),
 53-170 (1983), North-Holland, Amsterdam-New York.}

\ref{Hietarinta}{}{Direct methods for the search of the second invariant.
 Phys. Rep. {\bf 147}, 87-154 (1987).}

\ref{Krichever, I.M.:}{}{Elliptic solutions of the Kadomtsev-Petviashvili
equation
and integrable systems of particles. Funct. Anal. Appl. {\bf 14}, 282-289
(1981).}

\ref{See, for instance: Ruijsenaars, S.N.M., Schneider, H.:}{}{ A new
class of integrable
systems and its relation to solitons. Ann. Phys. (NY) {\bf 170}, 370-405
(1986);
Ruijsenaars, S.N.M.: Systems of Calogero-Moser type. {\it Proceedings} of
the 1994
Banff Summer School on Particles and Fields, CRM Proceedings and Lecture Notes
(in press),and the papers referred to there.}

\ref{Wojciechowsky, S.}{}{ Superintegrability of the 
Calogero-Moser system. Phys. Letters A {\bf 95}, 279-281 (1983).}

\end{document}